\documentclass[aps,twocolumn,showpacs]{revtex4}
\usepackage{graphicx}
\DeclareGraphicsExtensions{.eps,.ps,.eps.gz,.ps.gz,.eps.Z,{}}

\input{psfig.sty}

\begin{document}
\title{Ionization of Rydberg atoms embedded in an ultracold plasma}
\author{Nicolas Vanhaecke,$^{\dag }$ Daniel Comparat,$^{\dag }$$^{\ast }$ Duncan A. Tate,$^\ddag$ and Pierre Pillet$^{\dag }$}
\affiliation{$^{\dag }$Laboratoire Aim\'{e} Cotton, CNRS II, B\^{a}t. 505, Campus  d'Orsay, 91405 Orsay cedex, France}

\begin{abstract} We have studied the behavior of cold Rydberg atoms embedded in an ultracold plasma. We demonstrate that even
deeply bound Rydberg atoms are completely ionized in such an environment, due to electron collisions. Using a fast pulse extraction of the electrons from the plasma we found that
the number of excess positive charges, which is directly related to the electron temperature $T_e$,  is not strongly affected by the ionization of the Rydberg atoms.
Assuming a Michie-King equilibrium distribution, in analogy with globular star cluster dynamics, we estimate $T_e$.
 Without concluding on heating or cooling of the plasma by the Rydberg atoms,
  we discuss the range for changing the plasma temperature by adding Rydberg atoms.
\end{abstract}


\date \today

\maketitle

 One challenge in ultra-cold plasma physics is to reach the correlated regime where the Coulomb energy dominates the kinetic energy. One suggested way, unfortunately limited to non alkali ions, is to cool the plasma ions by lasers \cite{killian2003}. An alternative way might be to use the binding energy of Rydberg states as ``ice cubes'' to
cool the plasma. 
The physics of ultracold Rydberg gases \cite{anderson1998,mourachko1998} and ultracold plasmas \cite{kulin1999} formed by
laser excitation of a cold atomic sample have strong similarities. Indeed, Rydberg atom formation in an ultracold
plasma~\cite{killian2001} and spontaneous evolution of an ultracold Rydberg gas to plasma \cite{robinson2000} have been
demonstrated. The Rydberg ionization process starts with blackbody photoionization and initial electrons leave the
cloud region. A second phase occurs when the positive ion potential is deep
enough to trap subsequent electrons, which then collide with Rydberg atoms creating more electrons in an avalanche ionization
process~\cite{robicheaux2003,pohl2003,li2004}. However, other relevant processes have been proposed whose effects need to be
investigated, such as continuum lowering \cite{hahn2002}, and many-body effects or long-range interactions \cite{fioretti1999a}
that can lead to autoionization of Rydberg atom pairs~\cite{hahn2000}.

In this letter, we report the behavior of a mixture of an almost {\it neutral} ultracold plasma and a cold Rydberg atom sample. A
related experiment has been reported, but in the case of rubidium Rydberg atoms created in a purely {\it ionic}
plasma~\cite{feldbaum2002}. In this letter, we analyze the  fast avalanche ionization of deeply bound Rydberg states embedded in a quasi-neutral
plasma. We also study, with a theory based on analogy with globular star cluster dynamics, the evolution of the temperature of the plasma
when Rydberg atoms are added. 

The  cesium magneto-optical trap (MOT) apparatus has been described in a previous paper \cite{robinson2000}. Two dye lasers
pulses (Coumarin 500) that are focused to the cold atom cloud diameter excite atoms initially in the $6p_{3/2}$ state.  The time
origin of the experiment is the first laser (L$_1$) pulse, with typical energy $P_1=10\,\mu$J, which creates a quasi-neutral plasma of $N_i\approx 4 \times
10^5$ ions  with peak density $10^{10}$cm$^{-3}$. The second laser (L$_2$) pulse (ASE $<1\%$),  has a
$18\,$ns delay and  excites typically $4 \times 10^5$ Rydberg atoms. The Rydberg number fluctuates from pulse to pulse due to the
changing overlap of the  dye laser mode structure with the narrow $6p_{3/2}$ $\rightarrow$ Rydberg resonance
\cite{robinson2000}. However, the plasma created by the first laser, which is tuned just above the ioniziation limit, is affected
only by negligible laser intensity fluctuations. The MOT trapping lasers are turned off and a  resonant 852 nm diode laser
pulse excites the Cs
$6p_{3/2}$ state just before the dye laser pulses arrive. Because of the Doppler effect, this pulse ensures that all the $6p_{3/2}$ atoms are cold \cite{robinson2000}. We have verified that $6s$ atoms have no effect on our experimental results by pushing them away using this laser light pressure after the pulse dye laser excitation. 

\begin{figure}[hb]
\resizebox{0.5\textwidth}{0.3\textwidth}{
\rotatebox[origin=rB]{270}{
\includegraphics*[0cm,1.6cm][19cm,25.5cm]{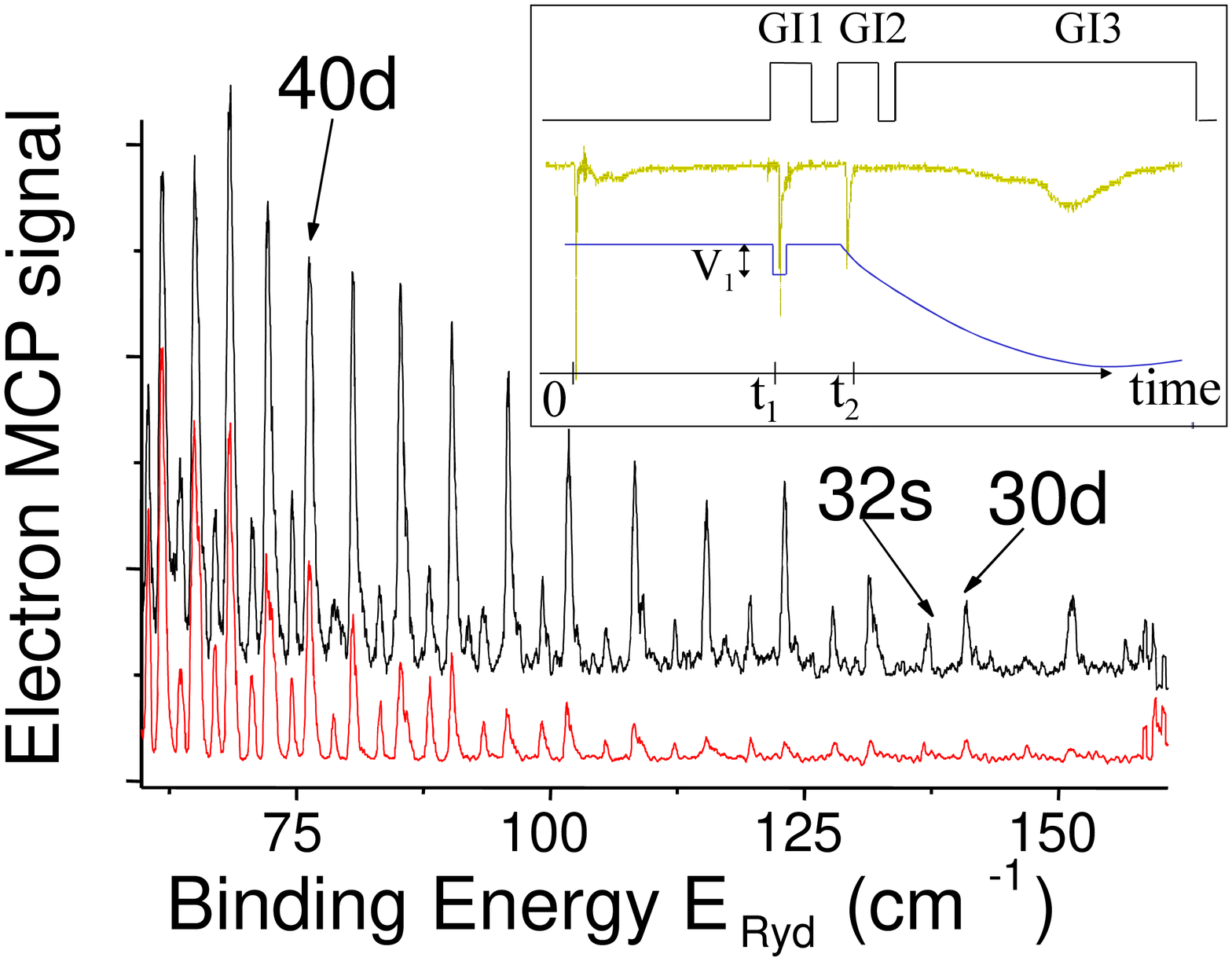}  } } 
\begin{center}
\caption{(Color online) Electron signal GI2 monitored at $t_1 = 4.7\,\mu$s after plasma creation when scanning the Rydberg laser. Upper curve:
plasma present. Lower curve: plasma absent. In the inset: lower trace: a schematic view of the voltage between the two grids, the
middle trace: MCP signal (the first peak is noise from the laser shot), the upper trace: integrator gates.}
\label{fig:scan}
\label{fig:signal}
\end{center}
\end{figure}

The first part of our analysis concerns the ionizing effect of the plasma on the Rydberg atoms. In order to study
the evolution of the cloud we applied, after a variable time $t_1 = 0-20\,\mu$s, a positive top hat voltage pulse (voltage $V_1$)
to one of the two parallel grids ($d=1.57\,$mm separation) that surround the cloud. This pulse is used to pull all the
electrons from the plasma (hereafter termed ``free'' electrons) toward a microchannel plate (MCP) detector, leaving no further
electrons trapped in the ion space charge in the situation where the laser L$_2$ is blocked. The
MCP signal is monitored using a gated integrator (GI1) (see inset in Fig.\ref{fig:signal}). The effect of the Rydberg atoms on
the plasma is monitored using a second voltage pulse applied on the opposite grid, at time $t_2$
typically
$700\,$ns delay from the first voltage pulse. The early part of this second pulse extracts  additional electrons
that are monitored using a second integrator (GI2). If the second voltage pulse is large enough it will also field ionize Rydberg states, leading to a signal on the MCP that is monitored using a third integrator, GI3. 
Fig.\ref{fig:scan} shows
the results, for $t_1=4\,\mu$s, versus the wavelength of L$_2$. L$_1$ is blocked by a mechanical shutter half of the time so that we obtained spectra with, and without, the plasma 
under otherwise similar conditions. The results are illustrated using the case of the $30d$ state. When only $30d$ atoms are created, almost no electrons are detected in GI1 or GI2.  However, when the plasma is created, some electrons appear in GI2 but the electron number in GI1 is
unchanged. This indicates that more free electrons are formed when the Rydberg laser is present. These electrons must come from the Rydberg atoms, indicating that the plasma
accelerates the ionization of the Rydberg atoms.

\begin{figure}[hb]
\resizebox{0.5\textwidth}{0.3\textwidth}{
\includegraphics*[2.9cm,1.3cm][27.0cm,20.0cm]{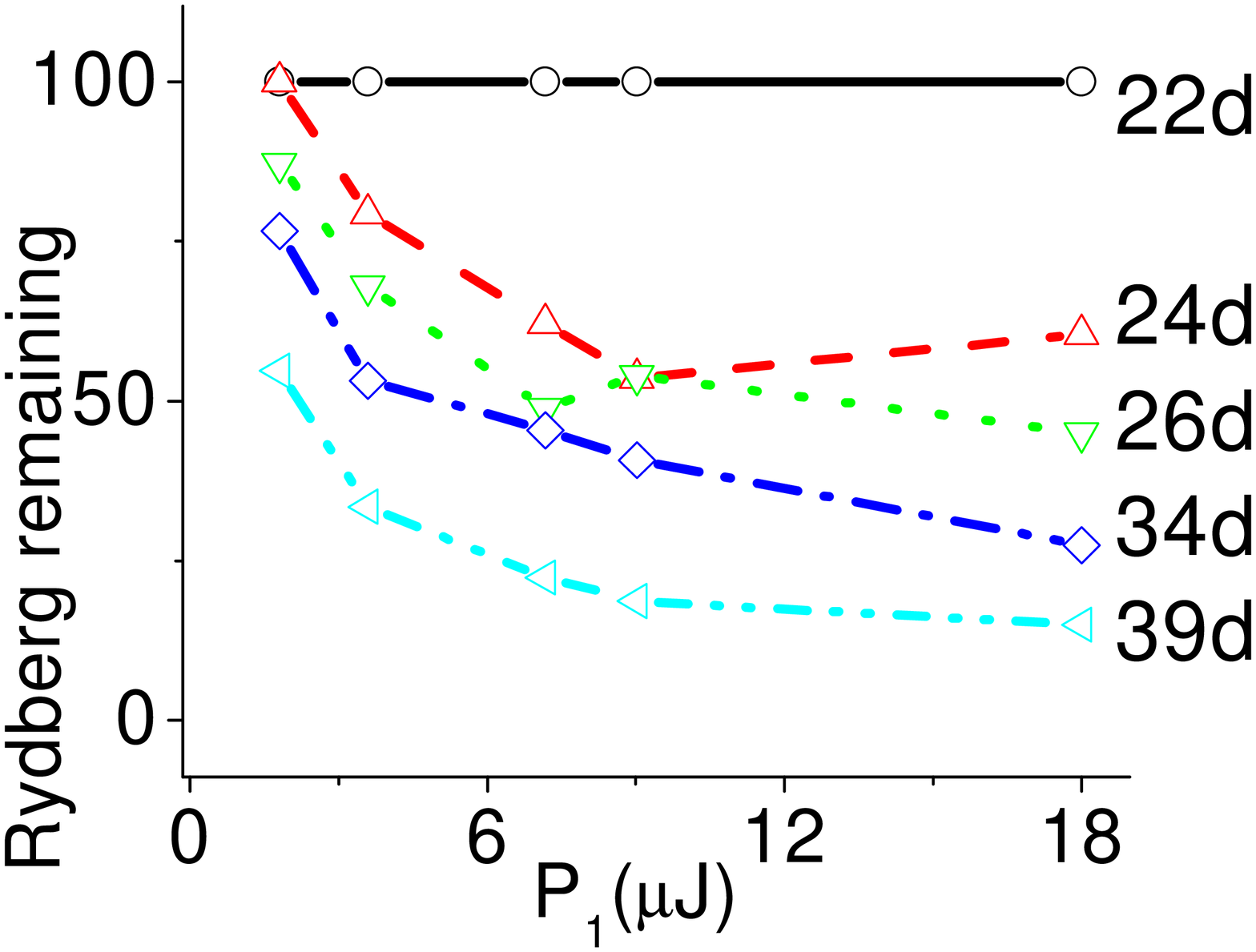}  } 
\begin{center}
\caption{(Color online) Rydberg ionization efficiency (in purcent) versus plasma laser power ($P_1=10\ \mu$J corresponds roughly to $4\times 10^5$ ions) at $t_1=1 \ \mu$s. The GI3 remaining
Rydberg atom signal is monitored.}
\label{fig:density}
\end{center}
\end{figure}

Looking at Rydberg ionization efficiency (GI3 signal) as a function of plasma density and binding energy of the Rydberg
state leads to the more quantitative results shown in Fig. \ref{fig:density}. The denser the plasma, the more efficient
the ionization. Furthermore, Rydberg atoms which do not ionize spontaneously can be ionized by the plasma. Long
plasma-Rydberg interaction times favor the Rydberg ionization process. For instance, we were able to ionize Rydberg states with
the plasma down to the dye limit of $n=19$ at delays of $10-20\,\mu$s. The maximum effective evolution time is
$20\,\mu$s, since after this delay plasma expansion leads to too small a density to have efficient ionization. It is also
worthwhile noting that similar results are obtained when the plasma is much smaller in size ($\sim 0.1 \ $mm diameter)
than the Rydberg sample ($\sim 0.6$ mm diameter). The Rydberg sample is still completely ionized (for long interaction time), presumably as a consequence of
the plasma expansion.

All of these observations agree with the model based on avalanche ionization due to collisions between plasma electrons and
Rydberg atoms. Some further observations help to understand the role of the other processes mentioned in the introduction. 
First, the effect of continuum lowering, which reflects the fact the zero of
energy of the isolated atoms is shifted by long-range Coulomb interaction with neighbors in the presence of the plasma, will
ionize only
$n \gtrsim
40$ at short delays for the highest density we could achieve \cite{hahn2002}.  This is in contradiction with the ionization of
the $19d$ state we observed after
$10\,\mu$s. Secondly, the efficiency of plasma-induced Rydberg ionization is strongly dependent on the plasma density but insensitive to Rydberg density, ruling out any possible effect of autoionization of Rydberg pairs on the ionization process. 
Third, we have checked, up to $100\,$cm$^{-1}$ above the ionization limit, that the ionization process does
not depend strongly on the plasma electron energy. Finally, if we remove all the free electrons just after plasma creation (at
$t_1=0.1\,\mu$s), preliminary results on small numbers of $30d$ Rydberg atoms show that after $10\,\mu$s, $0\%$  are ionized if
$P_1 < 15\,\mu$J. However, for $P_1 = 30\,\mu$J, $50\%$ are ionized, whereas $100\%$ are ionized if the electrons are not
removed at
$t_1=0.1\,\mu$s. At first glance, this seems incompatible with the scenario in which Rydberg ionization is caused by electron
collisions. However, for $P_1 = 30\,\mu$J when the free electrons are removed at $t=0.1\,\mu$s, the small number of electrons
subsequently formed by blackbody radiation or autoionizing atom pairs are trapped in the positive ion potential. We suggest
that in such an environment, they are able to ionize a large number of Rydberg atoms.  In addition, the creation rate for
autoionizing pairs may increase when electrons are present because they can stabilize the cloud expansion, and because of the
effect of attractive induced dipole-charge or dipole-dipole long range forces \cite{fioretti1999a}. More quantitative
experiments, to be published in
\cite{li2004}, that have been carried out at Colby College in collaboration with Laboratoire Aim\'e Cotton and with a group at University of Virginia have observed
redistribution to Rydberg states other than that initially populated by the laser.
This
seems to confirm the model in
which Rydberg ionization is caused primarily by electron collisions because
Rydberg redistribution is a necessary part of this process \cite{robicheaux2003}. 

\begin{figure}[hb]
\resizebox{0.5\textwidth}{0.3\textwidth}{
\includegraphics*[4.5cm,2.5cm][27.5cm,19.0cm]{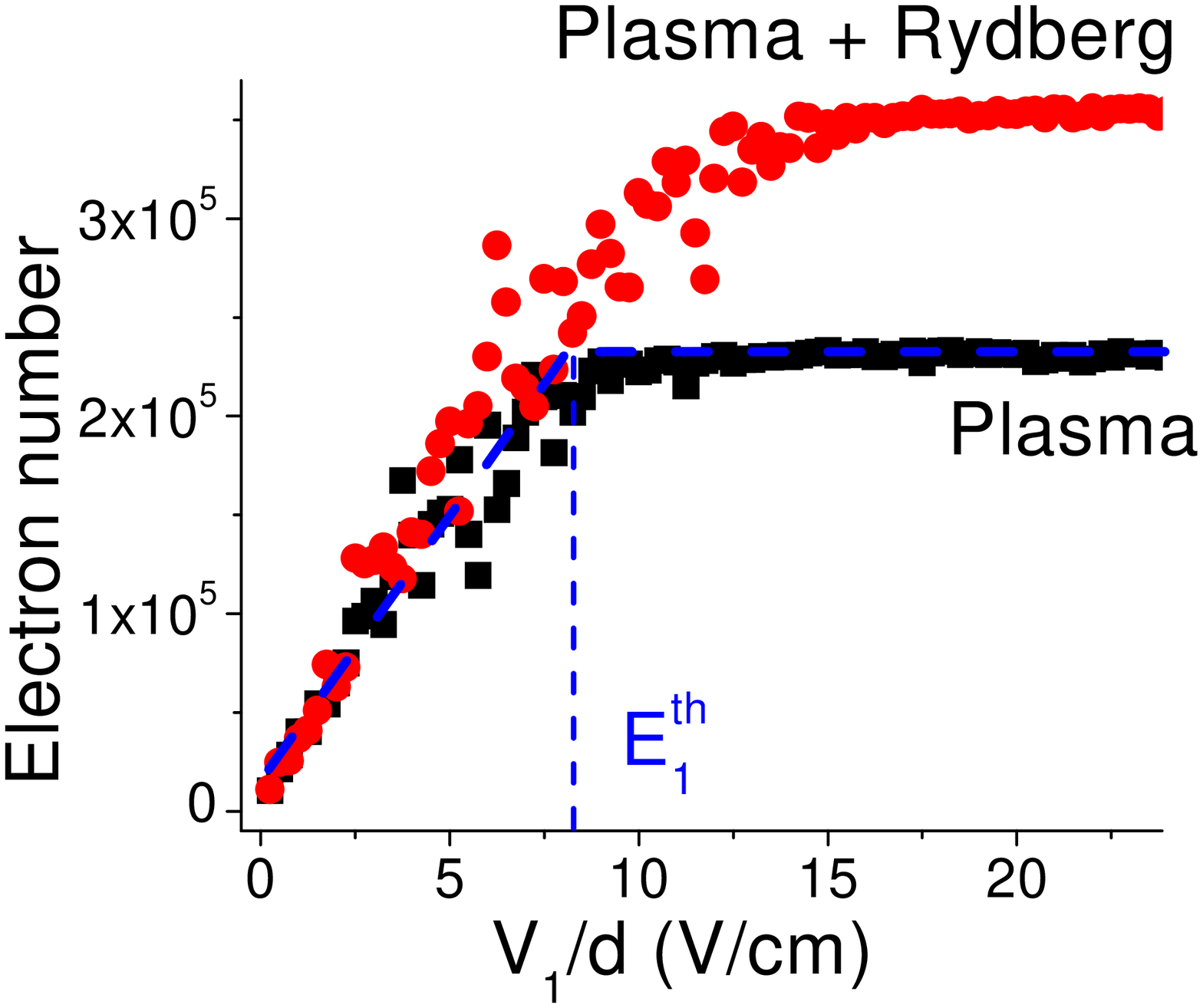}  } 
\begin{center}
\caption{(Color online) Number of electrons ejected (GI1) when varying the voltage $V_1$ for the plasma alone, and for $45d$ Rydberg atoms
embedded in the plasma after $t_1=1\,\mu$s of evolution time. The dashed lines are guide for the eye to interpret Eq.\ref{max_field_eq}. 
Each data have been calibrated to the multiple shots average electron number $\bar N_e$, i.e. the signal shown is $\bar N_e\times$GI1/(GI1+GI2).}
\label{fig:pulse}
\end{center}
\end{figure}

The second part of this letter is devoted to an important question concerning energy conservation during
the evolution of Rydberg atoms to plasma \cite{gallagher2003}. Electron collisions with very highly excited Rydberg atoms gradually decrease the binding energy of the Rydberg atoms until they ionize. On the other hand, electron collisions with lower-lying Rydberg states increase their binding energy (superelastic collisions). The interface between these two regimes occurs roughly
when the Rydberg binding energy equals few times the kinetic energy of the plasma electrons \cite{tkachev2001,robicheaux2003}. In the first scenario the Coulomb space charge is not affected and the electrons are probably cooled, so $T_e$ should decrease. However, in the second scenario the colliding electron is heated, or is ejected out of the plasma, and $T_e$ should increase.
To experimentally discriminate between these two mechanisms, we need to determine the electronic temperature of the plasma, $T_e$.
The basic idea to experimentally determine the temperature is to 
have instantaneous picture of the electron energy distribution
using the short voltage pulse $V_1$ to
extract electrons. In order to adequately accelerate the electrons when the pulse amplitude is small, its duration must be longer than $50\,$ns.  This is close to the electron thermalisation time and we probably also extract some rethermalized electrons.  A similar technique has been used in neutral atom Bose
Einstein Condensation experiments in a static external potential \cite{doyle1989}. Our case, electrons in the plasma, is more complex
due to the fact our potential $\phi$ (sum of the electronic, ionic and external potential) depends on the
number of particles trapped via Poisson's equation. Our experiment is similar to ``runaway electron'' experiments, except that
we have an inhomogeneous plasma, which leads to theoretical complexities \cite{kulsrud1973}. The number of electrons ejected  by
the voltage $V_1$ is plotted in Fig. \ref{fig:pulse} both for the plasma only, and for plasma plus Rydberg sample. Our lasers are not strongly focused so the cylindrical cloud is well approximated by a spherical gaussian symmetry for the ion density $n_i(r,t)=n_i^0
e^{-r^2/(2 \sigma^2(t))}$. For the data presented in Fig. \ref{fig:pulse} the sample is still gaussian (see
\cite{robicheaux2003}) with $\sigma=\sigma(t=0) \approx 250 \mu$m.

We first determine  $n_i^0=\frac{N_i}{(2\pi \sigma^2)^{3/2}}$
 using the threshold value $E_1^{\rm th}$ (see Fig. \ref{fig:pulse}) that is necessary to remove
all the free electrons. Indeed, assuming there are always a few electrons with zero velocity in the ion potential well,
the maximum electric field created by the ionic space charge is exactly
\begin{equation}
    E_1^{\rm th} = \frac{V_1^{\rm th}}{d} \approx 2.38 \frac{q_i}{4 \pi \varepsilon_0} n_i^0 \sqrt{2 \sigma^2}
    \label{max_field_eq}
\end{equation}
 Preliminary comparison with ion numbers coming from an ionic detection on the MCP detector are in agreement with this measurement.

We then determine the plasma state, immediately before we apply $V_1$ (time delay $t_1$), using the
Poisson-(Landau)-Fokker-Planck (FP) kinetic equation for electrons \cite{rosenbluth1957}.  
The cold and heavy ions lead to a negligible contribution in the electrons FP equation which is then is formally identical to the one governing globular cluster stars dynamics
\cite{spitzer1987,meylan1997}) except that, in the non collisional part, there is a repulsive electron-electron potential versus the attractive gravitational field.
The system reaches a quasi-equilibrium within a few times the electron-electron thermal momentum relaxation time $\tau_{\rm
ee}$, which is typically few tens of nanoseconds. The main result obtained with our analogy with globular clusters is then that the
phase-space density function $f$ for this quasi-equilibrium is close to a (Michie-)King type distribution \cite{binney1987}:
$$f\propto (e^{-E/k_B T_e}-e^{-E_t/k_B T_e})$$  where $k_B$ is Boltzman's constant, $E=q_e \phi+ \frac{1}{2} m_e v_e^2 $ and
$E_t$ is the potential energy at infinity \cite{comment2}. This defines what we call the uniform ($r$ independent) King's electronic
temperature $T_e$, which is not the same as the inhomogeneous velocity average temperature.
 Fitting the Monte Carlo simulations of Figs. 1 and 2 of Ref. \cite{robicheaux2003} confirms that such King's type distribution is a better choice than a Maxwell-Boltzmann distribution. 

The plasma temperature determination is then based on the solution of the self-consistent Poisson
equation:
  \begin{equation}
    \frac{1}{r} \frac{\partial^2}{\partial r^2} (r \eta_t(r,t)) = \frac{q_e}{k_B T_e(t)} \left[
    \frac{q_i n_i(r,t) + q_e n_e (r,t)}{\varepsilon_0}
    \right]
\end{equation}
where $\eta_t(r,t)=\frac{E_t(t)-q_e \phi(r,t)}{k_B T_e(t)}$ and
the electron density $n_e(r,t)$ is:
\begin{equation}
    n_e = \int_0^{+ \infty} f 4\pi v^2 {\rm d} v \propto e^{\eta_t}{\rm Erf}(\sqrt{\eta_t})-\frac{2}{\sqrt{\pi}}\left( \eta_t^{1/2}+\frac{2}{3}\eta_t^{3/2} \right)
    \label{density_ne_eq}
\end{equation}
where the proportionality factor is straightforwardly linked to the electron density at the cloud center, $n_e^0$, and to
$\eta (t) =\eta_t(t,r=0)$.
More precisely   $\sigma$ and $n_i^0$ are known
from Eq. \ref{max_field_eq} and we add the conditions: (i) $n_e(r)$ should be non-zero at infinity \cite{comment2},
because even for small $V_1$ values some electrons are removed, and (ii) the total number of calculated electrons
$N_e(t)=\int_0^{+
\infty} n_e(r,t) 4\pi r^2 {\rm d} r$ should reproduce the observed number $\bar N_e$. This determines the values for the two remaining
parameters
$n_e^0$ and $T_e$ in function of the unknown $\eta$ parameter.
The non analytical final results may be approximated by the following intuitive results:  the trapping depth $\eta k_B T_e$ has to be roughly equal to the trap depth,  calculated assuming a gaussian shape for ions and electrons: $\sqrt{2} q_e /(\sqrt{\pi} 4 \pi \varepsilon_0)$ times $(N_i-N_e)/ \sigma$ \cite{kulin1999}.
The experimental results (see Fig. \ref{fig:pulse})  indicates that $N_i-N_e$ is not strongly affected by the Rydberg ionization. 
This might be an indication that the superelastic collisions, which would lead to electron evaporation, are not a dominant process. This is also an indication that the sample is not strongly heated when Rydberg atoms are ionized. We cannot give a more definitive answer because the plasma temperature is mainly given by the ratio $N_i-N_e$ divided by the unknown parameter $\eta$.  Determination of $\eta$ from the shape of the curves in Fig. \ref{fig:pulse} is beyond the scope of this article, but this problem may probably be tackled using the collisionless Poisson-Vlasov ``violent relaxation''
theory.
Nevertheless,   $4<\eta<12$ is a reasonable choice, and for plasma alone, analysis of Fig. 2 of Ref. \cite{robicheaux2003} indicates that $\eta$ increases with time and that
$\eta \approx 10$ for our experiment with $t=1\,\mu$s (see also Fig. 10 of Ref. \cite{kuzmin2002b}). Without any simulation it seems difficult to know $\eta$ when Rydberg atoms are present in plasma.  Adding to the fact that $\sigma$ and $N_i-N_e$ are not strongly affected by Rydberg ionization. We could conclude that the temperature should not increase or decrease by more than a factor $5$ when Rydberg atoms are added to plasma. 

To conclude, we have studied the ionization of excited atoms in an ultracold plasma. We have observed the fast ionization of
Rydberg atoms embedded in the plasma. Deeply bound Rydberg atoms ($n=19$) are ionized using a dense plasma after some $10-20
\,\mu$s interaction time. The only mechanism able to explain a Rydberg ionization efficiency of close to 100\%  is
Rydberg-electron collisions from electrons trapped in
the ion space charge potential.
Nevertheless the role of long-range forces cannot be ruled out and should be checked with further experiments. Finally, in
analogy with globular cluster dynamics, we found that the plasma is always in a
Michie-King quasi-equilibrium distribution, this will be further discussed in a subsequent paper. $T_e$ is mainly given by $N_i-N_e$ which is experimentally determined by a forced fast electron extraction.
These preliminary experiments, with low a temperature plasma, indicate that Rydberg
atoms cannot drastically change the plasma temperature so this technique does not provide a path to reach the strongly coupled plasma regime.  
Measuring the electron evaporation rate with a static field or a voltage ramp are natural evolutions of the theory. 
Similar ideas have been published recently, \cite{roberts2004} but based on a Maxwellian distribution which we believe not to be appropriate in this system. 
We hope the analogy with globular clusters will be useful for the future of ultra-cold
plasma physics and will stimulate links with the astrophysics community.

The authors thanks 
Tom Gallagher for many helpful discussions.

This work is supported by the European Research Training Network ``QUAntum Complex Systems" and by Colby College.

$^{\dag}$Laboratoire Aim{\'{e}} Cotton is associated with Universit{\'{e}} Paris-Sud (website: www.lac.u-psud.fr) and is part of the Federation of research: "LUMAT".

$^{\ddag}$ Invited researcher by C.N.R.S. Permanent address: 
Department of Physics and Astronomy, Colby College, Waterville, Maine 04901, USA

$^\ast$Email address: Daniel.Comparat@lac.u-psud.fr

\addcontentsline{toc}{chapter}{Bibliographie}


\begin{thebibliography}{10}
\bibitem{comment2}
More generally $E_t$ is the truncated energy above which the particles are
  extracted from the sample. In our experiment the magnetic field gradient ($B'
  = 1.5\,$mT/cm) of the MOT ejects electrons for $r\gtrsim 12\sigma$. We use
  this value (and not infinity) to solve the Poisson's equation.
\end{thebibliography}
\end{document}